\documentclass[doublecol,a4paper]{epl2} 
\usepackage{hyperref,cuted}

\title{Unveiling causal activity of complex networks\footnote{The original version of this article was uploaded to the arXiv on March 17th, 2016 \cite{WilliamsGarcia2016}}}

\author{Rashid V. Williams-Garc\'{i}a\footnote{E-mail: rwgarcia@pitt.edu. Current affiliation: Departments of Neurobiology and Mathematics, University of Pittsburgh, Pittsburgh, Pennsylvania 15260, USA} \and John M. Beggs, \and Gerardo Ortiz}
\shortauthor{R. V. Williams-Garc\'{i}a \etal}
\shorttitle{Unveiling causal activity of complex networks}

\institute{Department of Physics, Indiana University, Bloomington, Indiana 47405, USA}

\abstract{We introduce a novel tool for analyzing complex network dynamics, allowing for cascades of causally-related events, which we call causal webs (c-webs), to be separated from other non-causally-related events. This tool shows that traditionally-conceived avalanches may contain mixtures of spatially-distinct but temporally-overlapping cascades of events, and dynamical disorder or noise. In contrast, c-webs separate these components, unveiling previously hidden features of the network and dynamics. We apply our method to mouse cortical data with resulting statistics which demonstrate for the first time that neuronal avalanches are not merely composed of causally-related events.}

\begin{document}

\maketitle

In systems consisting of many interacting elements, a variety of methods ({\it e.g.}, transfer entropy or Granger causality) are often used to reveal hidden dynamical causal links between them. This naturally leads to a complex networks description \cite{Gros2013}, raising interesting questions. For example, what fraction of the activity in such a network can be attributed to the hidden causal dynamics, and what fraction is produced by other processes, such as noise? Here we describe a new approach to this problem and demonstrate its utility on neural networks.

Over the past twenty years, there have been a number of theoretical \cite{Bak1996, BertschingerNatschlager2004, Chialvo2004, HaldemanBeggs2005, KinouchiCopelli2006, Levina2007, Kitzbichler2009, Chen2010, Larremore2011, MoraBialek2011, Pei2012, Manchanda2013, RybarschBornholdt2014} and experimental \cite{Worrell2002, BeggsPlenz2003, Pasquale2008, Poil2008, Shew2009, Ribeiro2010, Shew2011, Solovey2012, Priesemann2013, Shriki2013, Yu2013, deArcangelis2014, Shew2015} attempts to connect activity in living neural networks to critical avalanches like those seen in the Bak-Tang-Wiesenfeld (BTW) sandpile model \cite{Bak1987, ChristensenMoloney2005}. It has been hypothesized that homeostatic mechanisms might {\it tune} the brain, a complex neural network, towards optimality associated with a critical point \cite{NishimoriOrtiz2011} which separates ordered (``supercritical'') and disordered (``subcritical'') phases, where cascades of activity are amplified or damped, respectively \cite{HsuBeggs2006, Pearlmutter2009, Priesemann2013, deArcangelis2014, RybarschBornholdt2014}. In the BTW model, grains of ``sand'' are dropped one at a time at random lattice locations; sites which reach a threshold height topple their grains to their neighboring sites, potentially inducing further topplings, together forming an emergent cascade of events called an avalanche. Successive topplings are thus causally related, with each new toppling having been induced by another which happened before. The sandpile model eventually reaches a steady state in which the probability distribution of avalanche sizes follows a power law, a potential indicator of criticality. It is important to note that the grains are dropped at an infinitesimally slow rate such that the relaxation timescale, {\it i.e.}, the duration of the avalanches, is much shorter than the time between grain drops. This {\it separation of timescales} is essential to this concept of self-organized criticality (SOC) \cite{Bak1996}.

In real systems, however, this separation is not always achieved. As we will see, closer inspection of experimental neuronal avalanche data reveals potential conflicts with the SOC approach. For example, temporally distinct avalanches could be concatenated by sporadic events occurring between them, or two spatially distinct avalanches could be concatenated if they occurred synchronously. These confounding situations highlight the need for a method which clearly separates causally-related from independent activity. In our recent work, we developed a framework in which there is a \textbf{mixing of timescales}, as opposed to a separation of timescales \cite{WilliamsGarcia2014}. Using this framework, it was demonstrated that access to a critical point depends on the coupling of the concerned network to an external environment, resulting in a non-zero spontaneous activation probability $p_s$; the higher the probability, the further from criticality, and thus the further from optimality.

These spontaneous activation events could be caused by some unobserved influence, such as long-range innervation (in which case the network has been undersampled), activations which occurred prior to the start of an experiment, or the intrinsic properties of the network elements, such as those neurons which have a propensity to fire spontaneously or are tonically active, {\it e.g.}, as in the case of pacemaker cells or some inhibitory neurons \cite{FreundBuzsaki1996, Hu2014}. In neuronal cultures, spatiotemporal localization of cellular noise can lead to spontaneous avalanche production \cite{Orlandi2013}. Vanishing spontaneous activity (in combination with brief synaptic delays) is effectively equivalent to a separation of timescales, as described in SOC. Thus, achieving optimality by operating at a critical point may not be feasible for a living, open neural network, according to the quasicriticality hypothesis introduced in \cite{WilliamsGarcia2014}, although a relative optimality may still be achieved along a nonequilibrium Widom line \cite{WilliamsGarcia2014}. The accessibility of this relatively-optimal quasicritical region will then likely depend on the character of the environment and the fundamental properties of the neural network itself. Our ability to apply the nonequilibrium Widom line framework and test the quasicriticality hypothesis, however, hinges on the ability to identify spontaneous activity in a living neural network.

\begin{figure}
	\centering
	\onefigure[width = \columnwidth]{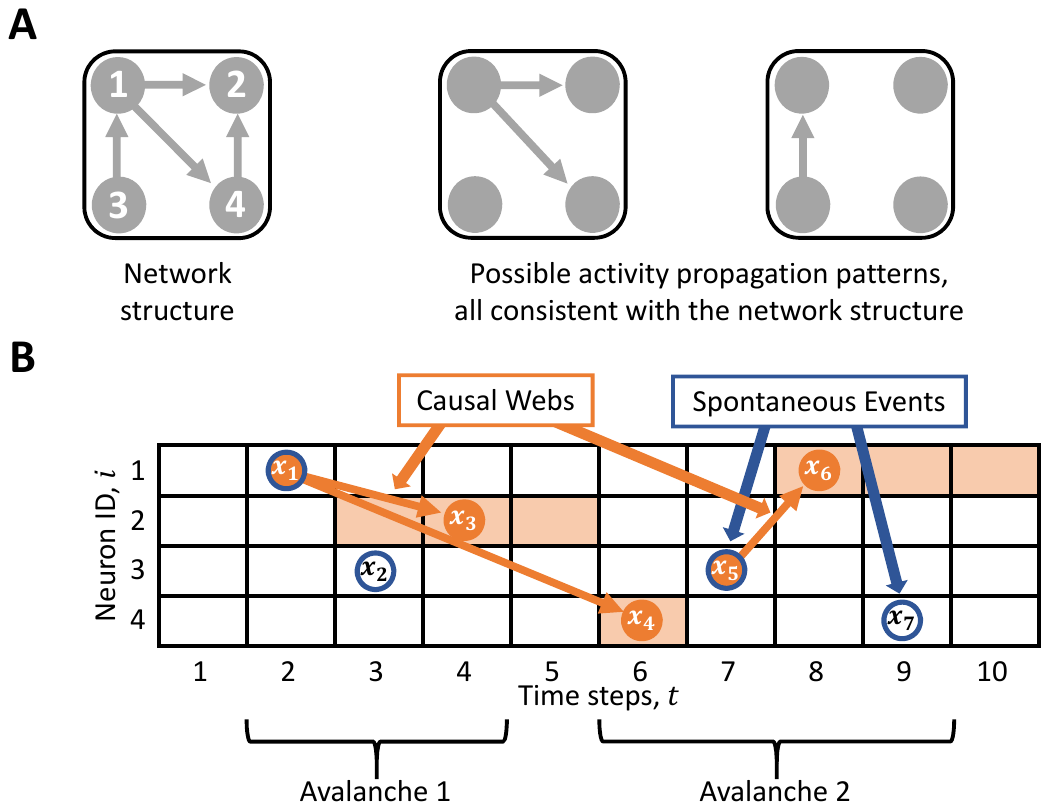}
	\caption{(Color online) Causal webs (c-webs) are distinct from avalanches in that they rely on network structure and connection delays. {\bf A.} This network produces a variety of spatiotemporal activity patterns (numbers correspond to neuron ID $i$). {\bf B.} A time raster showing the activity patterns in panel \textbf{A}; individual activations are labeled $x_\mu$. Whereas only two neuronal avalanches are detected, a richer structure is revealed when spontaneous events (blue annuli) are separated from c-webs (orange disks). Acceptance windows $W_{ij}(t)$ are shaded light-orange, where $i$ and $j$ correspond to different neuron indices, for example, $W_{12}(2)=[3,5]$. Notice that because of their contrasting definitions, c-webs and avalanches will generally have different statistical properties, e.g., c-webs may occur over a longer span of time due to connection delays, as shown here.}
	\label{Fig1}
\end{figure}

In this paper, we introduce a new method to disentangle spontaneous activity---which we define as activations occurring without an established causal link to a prior activation---from that which is causally-related and primarily governed by the network structure and dynamics. To this end, we next introduce the notion of {\bf causal webs} (or c-webs for short), as a new emergent cascade of correlated events, whose properties contrast and complement those of standard avalanches. Whereas the latter are defined as spatiotemporal patterns of activation spanning a number of adjacent time steps framed by time steps lacking activity, c-webs explicitly depend on the effective network and the temporal delays associated with the connections therein, thus accommodating the potential non-Markovian dynamics of complex networks. In the effective network, connections from source elements to target elements (which are analogous to presynaptic and postsynaptic neurons in physical neural networks) are established based on the predictive information that the activation of the source element provides of the activation of the target element; spontaneous events are those activations of element which occur in the absence of an established causal link to a prior activation of a source element. Knowledge of the network structure and delay information is key, as it allows to distinguish between different spatiotemporal patterns of activation in a way which is not possible with avalanches (see fig. \ref{Fig1}).

Let us formalize the concept of c-webs in the context of neural networks. We label individual events by $x=(i,t)$, representing the activation of neuron $i$ at time $t$, or following the notation used in \cite{WilliamsGarcia2014}, equivalently $z_i(t)=1$ ($z_i(t)=0$ meant quiescence and $z_i(t)>1$ corresponded to refractory states, which we do not consider to be activation events). We write the set of all events $A=\{x_\mu\}$, {\it e.g.}, in fig. \ref{Fig1}B, $A=\{x_1,x_2,x_3,x_4,x_5,x_6,x_7\}$. Formally, we define a c-web $C$ as a set $C=\{p_c\}_{c=1,|C|}$ ($|C|$ being the cardinality of the set $C$) of correlated ordered pairs $p_c=(x_\mu,x_\nu)_c$ of events ({\it i.e.}, spikes), which we call {\bf causal pairs}; quiescent and refractory neurons are not included in the set. The first and second entries, $x_\mu$ and $x_\nu$, of the $c$th causal pair represent causally-related source and target neuron events, respectively; each activation event can be associated to at most one c-web. (Despite causal relations being made in a pairwise fashion, we emphasize that this does not preclude multivariate interactions, as multiple pairings can be made to a single event.) In the following, we show how to determine those causal pairs.

A complete set of causal pairs $X$ is constructed by taking the Cartesian product of each event $x_\mu$ with its corresponding {\bf dynamic target neuron events} ${\cal U}(x_\mu)$, {\it i.e.}, $X= \bigcup_{x_\mu\in A} x_\mu\times {\cal U}(x_\mu)$, where ${\cal U}(x)\equiv {\cal U}(i,t)$ is the set given by
\begin{equation}
	{\cal U}(i,t) = \{(j,t^\prime)\; |\ j\in N(i) \mbox{ and } t^\prime\in W_{ij}(t)\}.
\end{equation}
$N(i)$ refers to the set of all target neurons $j$ of neuron $i$, and $W_{ij}(t)=[t+d_{ij}-\Delta_{ij},t+d_{ij}+\Delta_{ij}]$ is a predetermined dynamical {\bf acceptance window}: if a target neuron $j$ is active within the acceptance window, then a causal link between the activation events is inferred (see fig. \ref{Fig1}B). The lower bound of the acceptance window is adjusted such that it is greater than $t$. We write the set of distinct events in $X$ as $A(X)\subseteq A$.

Connection delays $d_{ij}$ associated with the connection from a source neuron $i$ to a target neuron $j$, are allowed to have some uncertainty $\Delta_{ij}$ due to variability in the target neuron spike timing. We will later present a method by which this information can be determined from data; for the moment, we assume it is given. In fig. \ref{Fig1}B, connection delays and their uncertainties are given for the connections in fig. \ref{Fig1}A: $d_{12}=2$, $d_{14}=4$, $d_{31}=2$, and $d_{42}=1$, with $\Delta_{12}=1$, $\Delta_{14}=0$, $\Delta_{31}=1$, and $\Delta_{42}=1$. This information can be used to determine causal pairs, {\it e.g.}, the event $x_1=(1,2)$ in fig. \ref{Fig1}B has ${\cal U}(x_1)=\{x_3,x_4\}$, resulting in the causal pairs $x_1\times {\cal U}(x_1)=\{(x_1,x_3),(x_1,x_4)\}$. The complete set of causal pairs for the {\bf spacetime graph} in fig. \ref{Fig1}B is $X=\{(x_1,x_3),(x_1,x_4),(x_5,x_6)\}$ and so $A(X)=\{x_1,x_3,x_4,x_5,x_6\}$. 

A causal web represents the connected components of a directed graph whose vertices and edges are $A(X)$ and $X$, respectively. The example in fig. \ref{Fig1}B thus has two c-webs, $C_1=\{(x_1,x_3),(x_1,x_4)\}$ and $C_2=\{(x_5,x_6)\}$. Note that spontaneous events initiate c-webs and may become part of ongoing c-webs; we call spontaneous events associated with a c-web $C$, its \textbf{roots} $r(C)$, e.g., $r(C_1)=\{x_1\}$ and $r(C_2)=\{x_5\}$. The {\bf size} $s(C)$ of a c-web is defined as the total number of distinct events within it. Defining that set as $A(C)$, the size $s(C)$ is then given by its cardinality: $s(C)=|A(C)|$. Note that $A(C)\subseteq A(X)$. For example, $A(C_1)=\{x_1,x_3,x_4\}$ and $A(C_2)=\{x_5,x_6\}$ in fig. \ref{Fig1}B, with $s(C_1)=3$ and $s(C_2)=2$, respectively.

The {\bf duration} $D(C)$ of a c-web $C$ can be defined in terms of its {\bf chord}. The chord of a c-web $K(C)$ is the sequence of distinct time steps for which there are events belonging to that c-web, arranged in ascending order in time, with no repeated elements. That is, $K(C) = (t_1,t_2,...,t_n)$, where $t_1$ and $t_n$ are the times of the first and last events, respectively. In contrast to the definition of duration for avalanches, the length of a c-web's chord is not equal to the c-web duration. Instead, we define the duration of a c-web as a measure of its chord plus one, {\it i.e.}, $D(C) = 1+\lambda(K(C))$, where $\lambda(K(C))=t_n-t_1$. The chords of the c-webs in fig. \ref{Fig1}B, for example, are $K(C_1)=(2,4,6)$ and $K(C_2)=(7,8)$, with durations $D(C_1)=5$ and $D(C_2)=2$.

Finally, we define the {\bf branching fraction} $b(C)$ of a c-web $C$ as the average number of target neuron events associated with each source neuron event:
\begin{equation}
	b(C) = \frac{1}{s(C)} \sum_{x_\alpha\in A(C)} \sum_{c=1}^{|C|} \delta_{x_\alpha,x_\mu},
\end{equation}
where $\delta$ is the Kronecker delta. The first sum is evaluated over all elements $x_\alpha$ of $A(C)$, while the second one is over all its causal pairs $p_c=(x_\mu,x_\nu)_c$, where $x_\mu$ represents the source neuron event of the pair $p_c\in C$. For example, in fig. \ref{Fig1}B, $b(C_1)=2/3$ and $b(C_2)=1/2$.

We performed tests of our method using simulations of the cortical branching model (CBM) \cite{WilliamsGarcia2014}. In the CBM, spontaneously activated nodes $i$ initiate cascades of activity which spread to neighboring nodes $j$ with activity transmission probabilities ({\it i.e.}, connection weights) $P_{ij}$ depending on their states $z_j\in\{0,1,2,...,\tau_r\}$, where $z_j=0$ corresponds to quiescence, $z_j=1$ to activation, and $z_j\in\{2,...,\tau_r\}$ correspond to refractory states. Following activation at time step $t$, $z_j(t)=1$, a node becomes refractory in the very next time step, $z_j(t+1)=z_j(t)+1$, and iterates until $z_j(t)=\tau_r$, after which the node returns to quiescence, $z_j(t+\tau_r)=0$. Only nodes which are quiescent at one time step can be activated in the following time step. There are no inhibitory nodes in the CBM.

\begin{figure}[!hb]
	\centering
	\onefigure[width = \columnwidth]{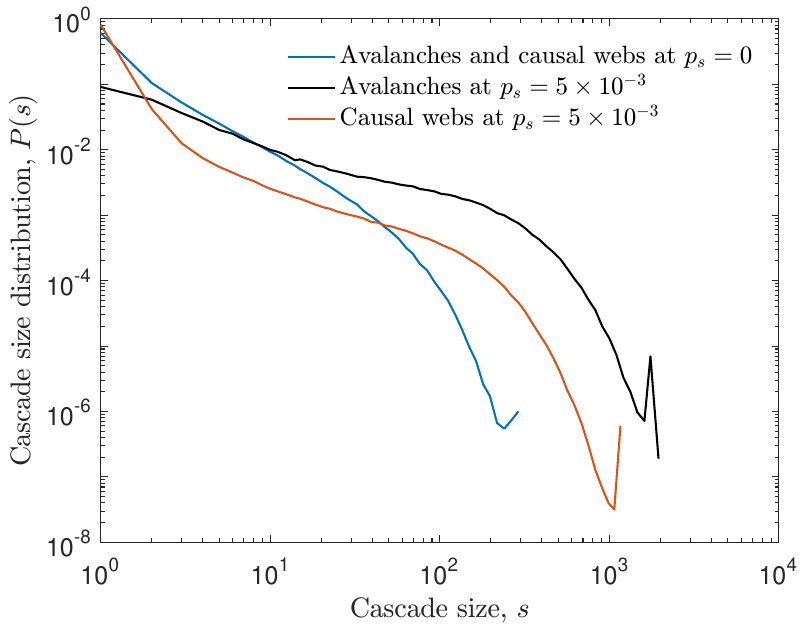}
	\caption{(Color online) Simulated avalanche and causal web size distributions coincide in the case of a separation of timescales (blue), but are substantially different under conditions corresponding to a mixing of timescales (orange/black). Simulations were performed in a data-inspired network of $N=243$ nodes.}
	\label{Fig2}
\end{figure}

As we have stated earlier, neuronal avalanches and c-webs should coincide as emergent cascades of correlated events in the limit $p_s\to0$ for all nodes and $d_{ij}=1$ for all pairs of nodes $(i,j)$. We simulated $10^6$ avalanches on a network of $N=243$ nodes, whose structure and connection weights were inspired by experimental data; all connection delays were set to a single time step, i.e., $d_{ij}=1$. To simulate the $p_s\to0$ limit (a separation of timescales), we initiated avalanches at single, random nodes, only starting a new avalanche when the previous one had finished; no spontaneous events or concurrent avalanches were allowed. The resulting avalanche and c-web size probability distributions were identical, as expected (see fig. \ref{Fig2}). In order to examine these distributions under conditions corresponding to a mixing of timescales, we allowed for nodes to become active spontaneously. Spontaneous activation probabilities for each node were drawn from a Gaussian distribution with mean and standard deviation of $5\times10^{-3}$; negative values were set to zero. As a result, the two distributions differ greatly; most notably, c-webs better capture the abundance of isolated spontaneous events ($s=1$ c-webs) than avalanches do.

In another test, we constructed a random network of $N=360$ nodes, each with an in-degree of $k_{in} = 3$, as in \cite{WilliamsGarcia2014}. The network was not strongly connected (i.e., it was reducible, thus contained subgraphs) and had a spectral radius (i.e., Perron-Frobenius eigenvalue) of $\kappa=0.23$. Connection delays (in time steps) were drawn from a uniform distribution of integers in a closed interval, $d_{ij} \in [1,16]$. Spontaneous activation probabilities for each node were drawn from a Gaussian distribution with mean and standard deviation of $10^{-4}$; negative values were again set to zero. The simulation was performed over $3.6\times10^6$ time steps and refractory periods of all nodes were set to a single time step (or, in the language of \cite{WilliamsGarcia2014}, $\tau_r=1$). Spontaneous events detected by our method were used to construct a new spontaneous activation probability distribution, which we compared with the initial distribution using a Kolmogorov-Smirnov (KS) test at a 5\% significance level: the distributions were in agreement with a p-value of $0.996$ \cite{James2006}. We note that as the overall connectivity of the network (which we quantify by $\kappa$, as in \cite{WilliamsGarcia2014}) is increased, spontaneous events become less prominent as c-webs begin to dominate the dynamics, leading to more driven activations (and, when $\tau_r>1$, refractory nodes), thus preventing spontaneous events: neural network dynamics present a fluctuating bottleneck to the influence of an environment.

To determine the c-webs, we have so far assumed knowledge of the network structure and delay information ({\it i.e.}, the $d_{ij}$ and $\Delta_{ij}$'s), however in practice, this information must be learned from experimental data. We now describe a method, based on delayed transfer entropy (TE) \cite{Ito2011,Wibral2013}, by which this information can be established from high temporal resolution multiunit timeseries data. The use of TE is not absolutely necessary; alternative measures of effective connectivity and causal effect, such as information flow, which also compute conditional probabilities directly from data, could in principle be used \cite{AyPolani2008, LizierProkopenko2010}. The notion of c-webs is completely independent of any one measure of effective connectivity.

For a particular pair of neurons $(i,j)$, TE is calculated at various delays $d$, peaking at the appropriate $d=d_{ij}$, with a width of $\Delta_{ij}$ \cite{Ito2011}. At a given $d$, the TE from neuron $i$ to neuron $j$, is given by
\begin{strip}
\begin{equation}\label{dTE}
	T_{i\to j}(d) = \sum_{{\bf z}_{i\to j}(d)} p(z_j(t-1),z_j(t),z_i(t-d))\log_2\left(\frac{p(z_j(t)|z_j(t-1),z_i(t-d))}{p(z_j(t)|z_j(t-1))}\right),
\end{equation}
\end{strip}

\begin{figure}
	\centering
	\onefigure[width=\columnwidth]{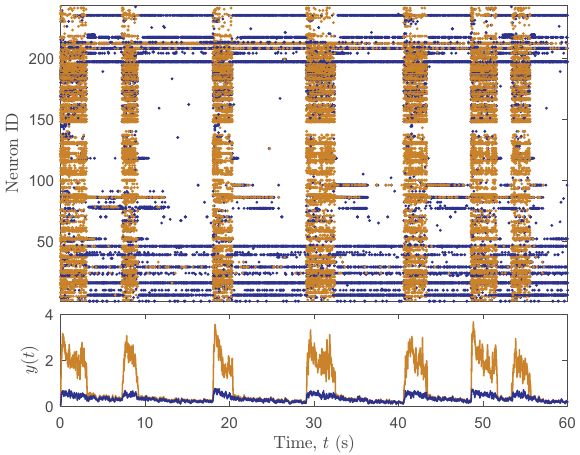}
	\caption{(Color online) One minute of neural network activity recorded from somatosensory cortex, after processing to separate spontaneous events (dark blue) from c-webs (orange) using the c-webs method. Note that tonically-active neurons mainly produce spontaneous events.}
	\label{Fig3}
\end{figure}

\begin{figure*}[ht]
	\centering
	\onefigure[width=\textwidth]{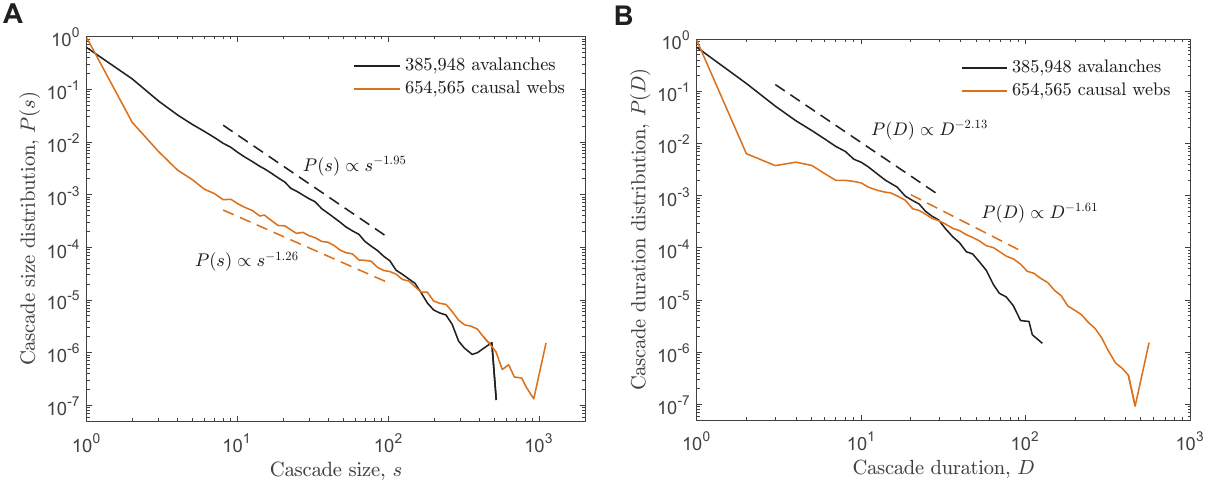}
	\caption{(Color online) Cascade size {\bf(A)} and duration {\bf(B)} probability distributions from simulations, which used the effective connectivity, delays, and times of spontaneous events extracted from \emph{in vitro} data using TE and c-webs analyses. Dashed lines represent guides to the eye over regions in which maximum likelihood estimates were performed.}
	\label{Fig4}
\end{figure*}

\noindent where ${\bf z}_{i\to j}(d) = \{z_j(t-1),z_j(t),z_i(t-d)\}$ indicates that the sum is performed over all possible configurations of the binary variables $z_j(t-1)$, $z_j(t)$, and $z_i(t-d)$. Joint and conditional probabilities in eq. (\ref{dTE}) are estimated from spike-sorted data, as in \cite{Ito2011}. For every pair of neurons $(i,j)$, $T_{i\to j}(d)$ is computed over a range of values $d \in [1,16]$; the peak value $T_{i\to j}(d_{ij})$ represents the putative connection from neuron $i$ to neuron $j$. The spike data is then randomly shuffled to establish a rejection threshold; TE values below this threshold are not considered to be significant and are set to zero \cite{BeggsPlenz2004, Ito2011}. The remaining TE values are then converted to activity transmission probabilities, as described in the supplemental materials of \cite{Friedman2012}, which assumes that spiking activity is Poisson-distributed. Note that the c-webs method could potentially be further improved by stochastically establishing causal pairs depending on those probabilities. Spurious connections, such as those due to common drive and transitive connections, are removed by considering delays of the significant connections; these results ``were valid over a wide range of values of the rejection threshold'' \cite{Nigam2016}. We use only the delays $d_{ij}$ of the significant connections to establish causal pairs.
\begin{figure*}[!hbtp]
	\centering
	\onefigure[width=\textwidth]{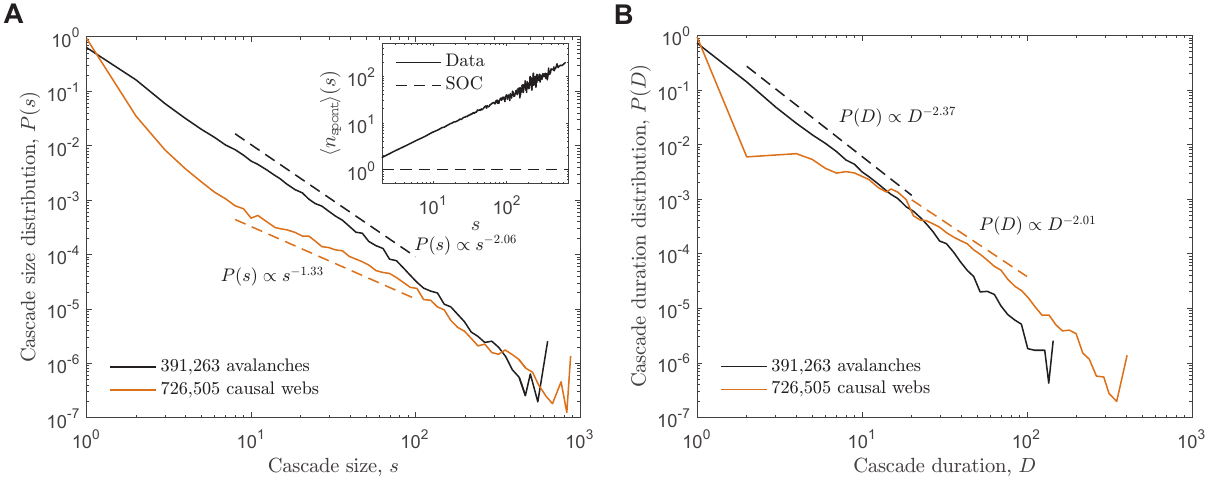}
	\caption{(Color online) Cascade size ({\bf A}) and duration ({\bf B}) probability distributions from \emph{in vitro} data. Avalanches (black) and c-webs (orange) exhibit different statistical properties due to a mixing of timescales (panel {\bf A} inset). Compare with the simulated predictions in fig. \ref{Fig4}; similarly, dashed lines represent guides to the eye.}
	\label{Fig5}
\end{figure*}

We next demonstrate the utility of our method when applied to experimental data (see fig. \ref{Fig3}). For our demonstration, we have used ten data sets from \cite{Ito2016}, which were collected {\it in vitro} from organotypic cultures of mouse somatosensory cortex using a 512-microelectrode array with a 60 $\mathrm{\mu m}$ electrode spacing and a 20 kHz sampling rate over hour-long recordings \cite{Litke2004, Ito2014}. Applying our method to a data set containing $N=243$ neurons, we extract spontaneous events (highlighted in dark blue) and c-webs (in orange) to illustrate their qualitative differences; note that spontaneous events may initiate or contribute to c-webs as in fig. \ref{Fig1}B. In fig. \ref{Fig3}, an activity time raster (top panel) is presented along with the corresponding timeseries of the activity (bottom panel), on which we have performed a moving average with a $\Delta t=100$ ms window: $y(t)=\sum_{t^\prime=0}^{\Delta t-1} x(t-t^\prime)/\Delta t$, where $x(t)=\sum_{i=1}^{N}\delta_{z_i(t),1}$ \cite{Oppenheim1989}.

We then performed simulations of the CBM using information extracted from the experimental data with TE and from applying the c-webs method, i.e., the activity transmission probabilities ({\it i.e.}, connection weights), connection delays, and spontaneous events. Using a data set which contained $N=435$ neurons, the connection weights were adjusted by a factor $\kappa$ (to manipulate the Perron-Frobenius eigenvalue of the corresponding adjacency matrix) over the range $[0.20,1.40]$ in steps of $\Delta\kappa=0.05$ and performed simulations at each value of $\kappa$---the raw data had a Perron-Frobenius eigenvalue of $\kappa\approx0.31$. Instead of stochastically initiating activity cascades using a fixed value of $p_s$ (as in \cite{WilliamsGarcia2014}), we used the spontaneous events identified from the data to initiate activity; as a result, each node exhibited a vastly different number of spontaneous activations. Refractory periods for each node were set to 1 time step and simulations were run over $3.6\times10^6$ time steps. The resulting avalanche (defined in 1 ms bins) and c-web size and duration probability distributions for simulations performed at $\kappa=0.80$ are plotted in fig. \ref{Fig4} using a logarithmic binning of $1.1$ (for a description of log-binning, refer to Appendix E of \cite{ChristensenMoloney2005}).

In fig. \ref{Fig5}, we plot size and duration probability distributions (again with logarithmic binning of $1.1$) using avalanches and c-webs identified directly from the data; as predicted in fig. \ref{Fig1}, maximum c-web durations were longer than those of avalanches in all ten data sets. Maximum likelihood estimation of the putative power law exponents was performed over the regions indicated by the dashed lines in fig. \ref{Fig5} using the methods described in \cite{Marshall2016}. The dashed lines simply serve as guides to the eye. This analysis produced putative exponents of $2.06$ and $1.33$ with log-likelihoods $-3.47$ and $-4.05$ for avalanche and c-web sizes, respectively (see fig. \ref{Fig5}A). The avalanche and c-web duration distributions (fig. \ref{Fig5}B) featured exponents of $2.37$ and $2.01$ with log-likelihoods of $-1.70$ and $-3.96$, respectively. Additionally, we note the emergence of isolated spontaneous activation events ($s=1$ c-webs) and dominance over larger c-webs, which does not fit the traditional picture of avalanche criticality. This demonstrates that while neuronal avalanches may exhibit apparent power-law scaling, thus suggesting potential underlying critical behavior, the corresponding c-web distributions do not; this is evident from both a qualitative and quantitative examination. Although we have not determined whether the data feature quasicritical dynamics, operating at or near the nonequilibrium Widom line, these results strongly suggest non-critical dynamics. To further illustrate how avalanches confound important dynamical information in the data, we plotted the average number of spontaneous events $\langle n_{\sf spont}\rangle(s)$, contained in avalanches as a function of their size $s$, and compared this to the expected result from a situation with a separation of timescales (which we verified using simulations), where SOC would be applicable (see fig. \ref{Fig5}A inset). In other words, our results put the application of the SOC framework to complex neural network dynamics into question.

In summary, we introduced here a novel spatiotemporal cascade of causally related activity, c-webs, to describe network dynamics in ways which contrast and complement conventional avalanches. This allows us to separate causally-related from unrelated events in complex network dynamics. Whereas avalanches strongly depend on the choice of temporal binning \cite{BeggsPlenz2003, Pasquale2008, Plenz2014}, c-webs only depend on the accuracy of the methods used to determine them---in this case, transfer entropy (TE). While TE provides good, model-free estimates of the information flow in a network ({\it i.e.}, the effective connectivity), it suffers from a number of limitations; for instance, TE requires establishing a suitable threshold for accepting putative connections and, in neural networks, it does not identify connections as excitatory or inhibitory \cite{Ito2011}. Additional experimental manipulations not performed here could possibly determine connection types when used in conjunction with TE (see, for example \cite{Orlandi2014}). Because of this, we have validated our c-webs method on an Izhikevich model network, with 80 excitatory regular spiking neurons and 20 inhibitory fast spiking neurons---using the sample network from \cite{Ito2011}---uniformly-distributed random synaptic delays between 1 ms and 20 ms for excitatory synapses, and 1-ms delays for inhibitory synapses. Individual neurons were stimulated with 10-Hz Poisson processes; activations triggered by Poisson inputs were marked as true spontaneous events. Under these conditions, the c-webs method was able to identify 71.3\% of the true spontaneous events, with a false positive rate of 18.2\%, demonstrating (1) the utility of the method when both excitatory and inhibitory neurons are present, and (2) that the ability to identify spontaneous events is limited by the ability of TE to identify true connections and their associated delays. Spurious connections detected by TE have been minimized here using recent methods \cite{Nigam2016}. We have used the cortical branching model (CBM) to compare statistical properties of c-webs with those of neuronal avalanches, finding that they entirely coincide in the absence of spontaneous events, {\it i.e.}, when $p_s=0$---a situation which corresponds with a separation of timescales---and differ significantly when $p_s>0$, i.e., when there is a mixing of timescales. Notably, the c-web size distribution does not appear to be scale-free---a result inconsistent with self-organized criticality (SOC). Indeed, application of our c-webs method to mouse cortical data demonstrated that neuronal avalanches are not merely composed of causally-related events (cf. fig. \ref{Fig5}). 

Potential further applications for c-webs in neural networks are numerous. For instance, c-webs could play a role in the characterization of different classes of neurons, network structures, and dynamical states. Because inhibitory neurons exhibit different firing patterns from excitatory neurons, {\it e.g.}, fast-spiking and tonic activation \cite{FreundBuzsaki1996, Hu2014}, distributions of spontaneous neuronal events may help identify inhibitory neurons, complementing established methods \cite{Bartho2004}. Hidden network cycles might be discovered by employing our approach in conjunction with population coupling methods \cite{Okun2015}. Additionally, c-webs may be used to distinguish recurrent from feed-forward network dynamics, by searching their structure for driven re-activations. Moreover, different dynamical network states and neurological disorders may be characterized by the prevalence of spontaneous events---a sort of signal-to-noise ratio. Finally, c-webs allow for a more careful examination of the quasicriticality hypothesis and practical application of the nonequilibrium Widom line framework \cite{WilliamsGarcia2014} to distinguish quasicriticality from criticality in living neural networks.

Similar applications could be envisioned for complex networks in general. For example, financial networks could be decomposed into agents that directly interact through exchanges as well as exogenous factors like weather or inflation. In climate research, identifying causal connections between different geographic regions is important for understanding the impact of localized events on global climate \cite{Runge2015}. In models of disease spreading, such as the SIRS model, c-webs could differentiate between sources of infection \cite{AndersonMay1979}. Such an approach is likely to be useful whenever considering interacting units, whether they are people in social networks, species in ecological webs, or protein molecules in a stochastic environment. A specific application in social media could involve the detection of Twitterbots and astroturfing \cite{Ratkiewicz2011}.

\acknowledgments
The authors would like to thank Hadi Hafizi, Emily B. Miller, Benjamin Nicholson, and Zachary C. Tosi for valuable discussions, as well as Shinya Ito, Alan M. Litke, and Fang-Chin Yeh for providing their {\it in vitro} data. R. V. Williams-Garc\'ia is presently supported by a National Institutes of Health Ruth L. Kirschstein National Research Service Award, T32 NS086749.

\end{document}